\documentclass[aps,showpacs,nofootinbib]{revtex4}

\usepackage{graphicx}

\begin{document}


\title{Is there any evidence for extra-dimensions or
quantum gravity effects from the delayed MeV-GeV photons in GRB
940217?}

\author{K. S. Cheng}
\email{hrspksc@hkucc.hku.hk} \affiliation{ Department of Physics,
The University of Hong Kong, Pokfulam Road, Hong Kong}

\author{T. Harko}
\email{harko@hkucc.hku.hk} \affiliation{ Department of Physics,
The University of Hong Kong, Pokfulam Road, Hong Kong}

\date{July 20, 2004}

\begin{abstract}
The discovery of X-ray afterglows of GRBs, and the identification
of host galaxies of GRBs, confirm the cosmological origin of GRBs.
However, the discovery of the delayed MeV-GeV photons in GRB940217
 imposes serious challenges for the standard emission
model of GRB. Although the delayed MeV-GeV photons might be
explained by some some radiation emission mechanisms, the mystery
of detecting an 18 GeV photon still remains unsolved. We suggest
that the detection of the 18-GeV photon $ \sim $4500 s after the
keV/MeV burst in GRB 940217 provides a strong evidence for the
existence of extra-dimensions and/or quantum gravity effects. The
delay scale of the 18-GeV photon leads to an estimation of the
fundamental energy scale, associated with the linear energy
dependence of the speed of light, of the order of $2.1\times
10^{15}$ GeV, which is consistent with the results obtained by
another independent analysis on the data of OSSE and BATSE.

\end{abstract}


\maketitle


\section{Introduction}

Delayed high-energy photon emissions from gamma-ray burst (GRBs)
sources can be considered a well-established fact. In 1994 Hurley
et al. \cite{Hu94} observed a particularly energetic burst, GRB
940217, with a duration of around 90 minutes. GRB 940217 is a very
strong burst, with a total fluence above $20$ keV of $(6.6\pm
2.7)\times 10^{-4}$ erg cm$^{-2}$ and a duration of $\sim $180 s
in the BATSE range. It has the third largest fluence of around 800
BATSE bursts up to the detection time of this burst. During the
period of low energy emission, i.e., in the first $\sim $180 s,
EGRET detected $10$ photons, with energies ranging from a few tens
of MeV to a few GeV, with the fluence in this range of $\sim
2\times 10^{-5}$ erg cm$^{-2}$. Most strikingly, after the
low-energy emission has ended, an additional $18$ high-energy
photons were recorded after $\sim $5400 s following this event,
including an $18$ GeV energy photon $\sim $4500 s and 36 photons
with 137 MeV energy. The fluence of the delayed emission was
$7\times 10^{-6}$ in the energy range $30$ MeV-$3$ GeV
\cite{Hu94}. For comparison, typical GRBs emit photons in the
energy range between a few keV and a few tens of MeV, and last a
few tens of seconds \cite {ChLu01}.

There are several possible astrophysical sources and mechanisms,
which could produce high energy gamma-ray photons. One such source
could be the Galactic diffuse gamma ray radiation, also detected
by EGRET \cite{Hu97}. The photon spectrum can be represented by a
broken power law, with a break energy at $1.9$ GeV. Above the
break energy the spectrum of the Galactic background radiation is
\cite{MaHa98}
\begin{equation}
F^{back}\left(E_{\gamma }\right)\approx 2.2\times
10^{-10}\left(\frac{E_{\gamma }}{1.9\text{ GeV}}\right)^{-3.1} .
\end{equation}

The number of background photons with energy greater than $10$ GeV
is given by
\begin{equation}
N_{\gamma }\left( E_{\gamma }>10\text{ GeV}\right) =\int_{10\text{GeV}%
}^{\infty }F_{\gamma }\left( E_{\gamma }\right) A_{eff}\Delta
tdE_{\gamma } ,
\end{equation}
where $A_{eff}$ is the effective, energy dependent area of the
gamma ray detector (700 cm$^{2}$ for $E_{\gamma }>10$ GeV in the
case of EGRET \cite{Hu97}) and $\Delta t$ is the duration of the
observation. For $\Delta t\approx 6000$ s we obtain $N_{\gamma
}\left( E_{\gamma }>10\text{ GeV}\right) \approx 6\times 10^{-5}$.
Therefore the $18$ GeV photon detected during GRB 940217 cannot be
of galactic origin.

An other possible source for the $18$ GeV photon in the GRB 940217
observation is the collision between the high energy (in the TeV
range) gamma-ray photons, produced in a GRB emission, and the
diffuse background of microwave or infra-red photons, leading to
electron-positron pair production. The pairs produced will scatter
off the background photons, and since the number density of
microwave background photons is significantly higher than that of
infra-red photons, many 2.7 K microwave radiation photons will be
boosted to a very high energy, in the (MeV-GeV) range
\cite{Wang2004}, \cite{ChCh96}. The spectrum of the scattered
cosmic microwave photons has a power law form \cite{Wang2004},
\cite{Dai2002}. The scattered photons will have a characteristic
energy $E_{\gamma }\leq (4/3)\gamma _{e}^{2}\left\langle
\varepsilon \right\rangle$, where $\left\langle \varepsilon
\right\rangle =2.7k_BT_{CMB}$ is the mean energy of the CMB
photons, with $k_B$ the Boltzmann constant and $T_{CMB}$ the
temperature of the cosmic microwave background. $\gamma_{e}$ is
the Lorentz factor of the secondary pairs resulted from a primary
photon with energy $\varepsilon _{\gamma }$. For $\gamma _{e}\leq
10^{6}$, taking into account that $\left\langle \varepsilon
\right\rangle \approx 10^{-4}$, it follows that the energy of the
scattered photons is of the order of $E_{\gamma }\leq 200$ MeV.
Therefore the $18$ GeV photon from GRB 940217 cannot be a
reprocessed primary TeV photon.

The redshift at which GRB 940217 was produced has been estimated
in \cite{SaSt98}, by considering the opacity of the intergalactic
space to high energy gamma-rays. This depends upon the number
density of soft target photons. $\gamma $-rays at above $20$ GeV
will be attenuated if they are emitted at a redshift of $z\sim 3$.
Therefore the highest energy photon in this burst is constrained
to have originated at a redshift less than $z\sim 2$
\cite{SaSt98}.

Several physical mechanisms, which could explain the time delay
between MeV-GeV photons in GRB 940217, have been proposed in the
past years. The existence of an inter-galactic magnetic field
(IGMF) could give a natural explanation of this delay \cite{Pl95},
\cite{WaCo96}. However, the IGMF has not been detected so far,
Faraday rotation measures setting a limit of $\sim 10^{-9}$ G for
a field with 1 Mpc correlation length. Theoretical calculations
show that these fields could be as low as $10^{-20}$ or even
$10^{-29}$ G \cite{Kr94}. Some studies suggest that the IGMF could
be generated from a much weaker primordial magnetic field, $\sim
10^{-20}$ G, due to turbulence induced in the formation of large
scale structure in the universe \cite{Ku95}. Therefore no
convincing evidence for the existence of IGMFs exists. It has also
been suggested that the delay could be the result of the collapse
of a compact object \cite{MeRe94} or coalescence of a compact
binary \cite{Ka94}.

The time delay between soft emission and high-energy emission is
also suggested by some GRB models. Photon-pair electromagnetic
cascades can produce delayed MeV-GeV photons \cite{ChCh96}. In
\cite{ChMa95} autocorrelation was used to determine the duration
of substructures in different energy bands of selected bursts.
Electron inverse Compton emission from afterglow shocks could
produce a delayed GeV component \cite{ZhMe02}. The information
provided by the delay time could give constraints on models for
TeV gamma rays and could differentiate between mechanisms causing
the time delay \cite{Wang2004}. However, all these models
mentioned above cannot give a convincing explanation for the
detection of the 18-GeV delayed photon in GRB 940217.

It is the purpose of the present paper to suggest an other
possibility for the explanation of the GeV-MeV photon delay in GRB
940217, namely, the possibility that this effect is due to the
presence of extra space-time dimensions and/or to related quantum
gravitational effects. The delay scale of the 18-GeV photon leads
to an estimation of the fundamental energy scale, associated with
the linear energy dependence of the speed of light, of the order
of $2.1\times 10^{15}$ GeV, which is consistent with the results
obtained by another independent analysis on the data of OSSE and
BATSE \cite{El03}.

The present paper is organized as follows. In Section II we
briefly review the possible effects of extra-dimensions on the
propagation of gamma photons emitted in GRB explosions. The time
delay for photons in multi-dimensional universes is considered in
Section III. In Section IV we apply the derived results for the
case of the 18 GeV photon in GRB 940217.

\section{Could extra-dimensions be detected in astrophysical
processes?}

One of the most challenging issues of modern physics is the
existence of the extra-dimensions, idea proposed originally by
Kaluza in 1921 \cite{Kl19} and developed by Klein in 1926
\cite{Kl26}. Multi-dimensional geometries are the natural
framework for the modern string/M theories \cite{Wi96} or brane
models \cite{HoWi96}. String and Yang-Mills type models also
provide a natural and self-consistent explanation for the possible
variation of the fundamental constants \cite{Fo79}, as initially
suggested by Dirac \cite {Di38}. Hence the problem of the
extra-dimensions of the space-time continuum is closely related to
the problem of the variations of the fundamental constants, like,
for example, the fine structure constant or the speed of light
(for a recent review of the experimental and theoretical studies
and the present status of these fields see \cite{Uz03} and \cite
{Ma03}). Most of the theories with extra-dimensions contain a
built-in mechanism, which allows the variation of the fundamental
constants. Within the multi-dimensional approach the physical
interactions are described by a theory formulated in $4+D$
dimensions, and the conventional four-dimensional theory appears
as a result of a process of dimensional reduction. Couplings in
four dimensions are determined by a set of constants of the
multidimensional theory and the size $A$ of the space of
extra-dimensions. The multi-dimensional constants are assumed to
be genuinely fundamental and, consequently, they do not vary with
time. On the other hand it is natural to assume that in an
astrophysical or cosmological context $A$ varies with time,
similarly to the scale factor $a$ of our four-dimensional
Universe. But this leads in four dimensions to the time variation
of the fundamental constants, like fine structure constant $\alpha
$ or the gravitational coupling $G$. Moreover, since their time
dependence is given by the same factor $A$, the time variations of
$\alpha $ and $G$ could be correlated \cite{Lo03}.

The search for a unification of quantum mechanics and gravity is
likely to require a drastic modification of the present day
deterministic representation of the space-time properties. There
is at present no complete mathematical model for quantum gravity,
and no one of the many different models proposed so far can give a
satisfactory description of the physics on characteristic scales
near the Planck length $l_{P}$. However, in several of the
approaches used to find a theory of quantum gravity the vacuum can
acquire non-trivial optical properties, because of the
gravitational recoil effects induced by the motion of the
energetic particles. The recoil effects may induce a non-trivial
refractive index, with photons at different energies travelling at
different velocities \cite{El00c}. Photon
polarization in a quantum space-time may also induce birefringence \cite{Ga99}%
, while stochastic effects in the vacuum could give rise to an
energy
dependent diffusive spread in the velocities of different photons \cite{El00}%
.

Therefore a large class of physical models, incorporating quantum
gravitation and/or multi-dimensional field theories predict that
the propagation of particle in vacuum is modified, due to the
supplementary effects induced by the modification of the standard
general relativity. In particular, the possible violation of the
Lorentz invariance or the existence of extra-dimensions can be
investigated by studying the propagation of high energy photons
emitted by distant astrophysical sources \cite{Am98}.

The confirmation that at least some gamma-ray bursts (GRBs) are
indeed at cosmological distances raises the possibility that
observations of these could provide interesting constraints on the
fundamental laws of physics. The fine-scale time structure and
hard spectra of GRB emissions are very sensitive to the possible
dispersion of electromagnetic waves in vacuo, with velocity
differences $\Delta u\sim E/E_{QG}$, as suggested in some
approaches to quantum gravity. GRB measurements might be sensitive
to a dispersion scale $E_{QG}$ comparable to the Planck energy
scale $E_{P}\sim 10^{19}$ GeV, sufficient to test some of these
theories \cite{Am98}. Hence the study of short-duration photon
bursts propagating over cosmological distances is the most
promising way to probe the quantum gravitational and/or the
extra-dimensional effects. The modification of the group velocity
of the photons by the quantum effects would affect the
simultaneity of the arrival times of photons with different
energies. Thus, given a distant, transient source of photons, one
could measure the differences in the arrival times of sharp
transitions in the signals in different energy bands. A key issue
in such a probe is to distinguish the effects of the
quantum-gravity/multi-dimensional medium from any intrinsic delay
in the emission of particles of different energies by the source.
The quantum-gravity effects should increase with the redshift of
the source, whereas source effects would be independent of the
redshift in the absence of any cosmological evolution effects.
Therefore it is preferable to use transient sources with a known
spread in the redshift $z$. The best way to probe the time lags
that might arise from quantum gravity effects is to use GRBs with
known redshifts, which range up to $z\sim 5$.

Data on GRBs may be used to set limits on variations in the
velocity of light due to quantum gravitational effects. This has
been done, by using BATSE and OSSE observations of the GRBs that
have recently been identified optically, and for which precise
redshifts are available, in \cite{Ma00}.  A regression analysis
can be performed to look for an energy-dependent effect that
should correlate with redshift. The analysis of GRBs data yield a
limit $M_{QG}\sim 10^{15}$ GeV for the quantum gravity scale. The
study of the the times of flight of radiation from gamma-ray
bursts with known redshifts has been considerably improved by
using a wavelet shrinkage procedure for noise removal and a
wavelet `zoom' technique to define with high accuracy the timings
of sharp transitions in GRB light curves \cite{El03}. This
procedure optimizes the sensitivity of experimental probes of any
energy dependence of the velocity of light. These wavelet
techniques have been applied to $64$ ms and TTE data from BATSE,
and OSSE data. A search for time lags between sharp transients in
GRBs light curves in different energy bands yields the lower limit
$M_{QG}\geq 6.9\times 10^{15}$GeV on the quantum-gravity scale in
any model with a linear dependence of the velocity of light,
$c\sim E/M_{QG}$.

\section{Photon delay in multi-dimensional universes}

The general expressions for the time delay of photons of different
energies in the framework of multi-dimensional cosmological models
was considered in \cite{HaCh04}. In models with compactified
extra-dimensions (Kaluza-Klein type models), the main source of
the photon time delay is the time variation of the electromagnetic
coupling, due to dimensional reduction, which induces an
energy-dependence of the speed of light. A similar relation
between the fine structure constant and the multi-dimensional
gauge couplings also appears in models with large
(non-compactified) extra-dimensions. For photons of energies
around 1 TeV, propagating on cosmological distances in an
expanding Universe, the time delay could range from a few seconds
in the case of Kaluza-Klein models to a few days for models with
large extra-dimensions. As a consequence of the multi-dimensional
effects, the intrinsic time profiles at the emitter rest frame
differ from the detected time profiles. The formalism developed in
\cite{HaCh04} also allows the transformation of the predicted
light curves of various energy ranges of the emitter into the
frame of the observer, for comparison with observations. Therefore
the study of energy and redshift dependence of the time delay of
photons, emitted by astrophysical sources at cosmological
distances, could discriminate between the different
multi-dimensional models and/or quantum gravity effects.

Let's consider the propagation of gamma-rays from GRBs in the
Kaluza-Klein type models. For simplicity we restrict our
discussion to the five-dimensional case. Hence we assume a flat
Friedmann-Robertson-Walker type background metric of the form
\begin{equation}
ds^{2}=c^{2}dt^{2}-a^{2}\left( t\right) \left[ dr^{2}+r^{2}\left(
d\theta ^{2}+\sin ^{2}\theta d\varphi ^{2}\right) \right]
+\varepsilon \Phi ^{2}\left( t\right) dv^{2},
\end{equation}
where $a(t)$ is the scale factor of the Universe, $\varepsilon =\pm 1$ and $%
\Phi (t)$ is the scale factor of the fifth dimension, denoted by
$v$. We also assume that the time variation of the fine-structure
constant is entirely due to the change in the speed of light $c$.
Therefore we neglect any possible time variation of the electric
charge or Planck's constant. Then the time variation of the speed
of light can be related to the size of the fifth dimension by
means of the general equation
\begin{equation}
\frac{\Delta \dot{c}}{\Delta c}=\beta \frac{\dot{\Phi }}{\Phi } ,
\end{equation}
where $\beta =1$ in the case of the Einstein-Yang-Mills model and
$\beta =3$ for the case of the pure Einstein gravity in five
dimensions \cite{Lo03}. Hence for the variation of the speed of
light we obtain
\begin{equation}
c=c_{0}\left(1+\Phi ^{\beta }\right),
\end{equation}
where $c_{0}$ is the four-dimensional speed of light.

In order to find a simple, and directly testable relation between
the radius of the extra-dimension and the energy of the photon, we
shall assume, following the initial proposal in \cite{Ma90} that
the mass of a body (and the associated energy) corresponds to the
length of a ''line segment'' of the fifth subspace. In a more
general formulation, we shall assume that the variables parameters
$c$, $G$ and the photon energy $E=h\nu $ are related to the metric
tensor component of the fifth dimension by means of the equation
 \cite{HaCh04}, \cite{MaHa99}
\begin{equation}\label{mass}
\frac{G\left( t\right) E}{c^{4}}=\varepsilon\int_{v^{0}}^{v}\sqrt{\left| g_{44}\right| }%
dv=\varepsilon\int_{v^{0}}^{v}\Phi dv.
\end{equation}

Therefore the energy-dependence of the speed of light of the
photon due to the presence of an extra-dimension is given by
\begin{equation}
c=c_{0}\left[1+\varepsilon\left( \frac{E}{E_{K}}\right) ^{\beta
}\right], \label{eq7}
\end{equation}
where we denoted $E_{K}=c^{4}\Delta v/G$, with $\Delta v=v-v^{0}$
describing the variation of the size of the fifth dimension
between the moments of the emission and detection of a photon.

In the case of isotropic homogeneous cosmological models with
large non-compact extra-dimensions \cite{RaSu99} there is a
non-zero contribution in the four-dimensional space-time (the
brane) from the $5$-dimensional Weyl tensor from the bulk,
expressed by a scalar term $U$, called dark radiation \cite{Ha03}.
The ``dark radiation'' term is a pure bulk (five dimensional)
effect, therefore we cannot determine its expression without
solving the complete system of field equations in $5$ dimensions.
In the case of a Friedmann-Roberston-Walker type cosmological
model the expression of the dark radiation is $U=U_{0}/a^{4}$
\cite{Ha03}, with $U_{0}$ an arbitrary constant of integration.

For light propagating from cosmological distances the differential
relation between time and redshift is \cite{El03},
\begin{equation}
dt=-\frac{H_{0}^{-1}dz}{\left( 1+z\right) g(z)} ,
\end{equation}
where
\begin{equation}
g\left( z\right) =\sqrt{\Omega _{\Lambda }+\Omega _{M}\left(
1+z\right) ^{3}+\Omega _{U}\left( 1+z\right) ^{4}},
\end{equation}
and $H_{0}=72$ km s$^{-1}$ Mpc$^{-1}$, $\Omega _{M}\approx 0.3$ and $%
\Omega _{\Lambda }\approx 0.7$  are the Hubble constant, the mass
density parameter and the dark energy parameter, respectively
\cite{Fr01}, \cite{Pe03}. $\Omega _{U}$ is the dark radiation
parameter.

By taking into account the expressions for $\Delta c$ we obtain
the following general equation describing the time delay of two
photons:
\begin{equation}\label{1}
\Delta t=H_{0}^{-1}f^{(\beta )}\left( E_{1},E_{2}\right) \int_{0}^{z}\frac{%
\left( 1+z\right) ^{\beta -1}}{g\left( z\right) }dz,
\end{equation}
where the functions $f^{(\beta )}\left( E_{1},E_{2}\right)$,
$\beta =1,2,3$, describe the different physical models
incorporating extra-dimensional and/or quantum gravitational
effects. For $\beta =1$ we have the linear model, with
\begin{equation}
f^{(1)}\left( E_{1},E_{2}\right)=\frac{\Delta E}{E_K^{(1)}},
\end{equation}
where we denoted $\Delta E=E_{1}-E_{2}$. The linear model
corresponds, from the point of view of the extra-dimensional
interpretation, to an Einstein-Yang-Mills type model
\cite{HaCh04}. A linear energy dependence of the difference of the
photon velocities has also been considered in \cite{El03} as a
result of the dispersion-relation analysis of the Maxwell
equations in the non-trivial background metric perturbed by the
recoil of a massive space-time defect during the scattering of a
low energy photon or neutrino. A quadratic model of the form
\begin{equation}
f^{(2)}\left( E_{1},E_{2}\right)=\left(\frac{\Delta
E}{E_K^{(2)}}\right)^2,
\end{equation}
can also be considered in quantum gravitational models, in which
selection rules, such as rotational invariance, forbids first
order terms \cite{El03}. The function
\begin{equation}
f^{(3)}\left(
E_{1},E_{2}\right)=\frac{\left(E_{1}^{3}-E_{2}^{3}\right)}{E_K^{(3)}},
\end{equation}
corresponding to $\beta =3$, describes the effect of a pure
five-dimensional gravity on the propagation of light in
four-dimensions \cite{HaCh04}. In the above equations we denoted
by $E_K^{(\beta )}$, $\beta =1,2,3$ the energy scales associated
with the different types of extra-dimensional and/or quantum
gravity mechanisms.

In the case of non-compactified extra-dimensions, since the fifth
dimension is large, the scale factor $\Phi $ can also be a
function of $v$, and hence an explicit knowledge of the
$v$-dependence of $\Phi $ is needed in order to derive the speed
of light- photon energy dependence. However, taking into account
that there is a linear dependence of $\alpha $ on the scale of the
fifth dimension, we can assume again that the time delay between
two different energy photons emitted during a gamma ray burst is
given by the linear model \cite{HaCh04}.

\section{Discussions and final remarks}

By using Eq. (\ref{1}) we can evaluate the extra-dimensional and
/or quantum gravity energy scale, which follows from the time
delay of the $18$ GeV photon in GRB 940217. Assuming that the
photon originated at $z\sim 2$ and taking $\Delta t\approx
4500\pm800$ s, where we have also included the uncertainty in the
moment of photon emission during the burst, we obtain for the
linear energy scale the value
\begin{equation}
E_K^{(1)}\approx \left(1.75-2.51\right)\times 10^{15} {\rm GeV}.
\end{equation}

This value is very close to the value $E_K^{(1)}\approx 6.9\times
10^{15}$ GeV obtained by using a wavelet technique analysis of
BATSE and OSSE data \cite{El03}. For the quadratic model energy
scale we have
\begin{equation}
E_K^{(2)}\approx \left(1.77-2.12\right)\times 10^{8} {\rm GeV}.
\end{equation}

In the case of quadratic quantum gravity corrections the
corresponding energy scale, derived by using BATSE and OSSE data,
is $E_K^{(2)}\geq 2.9\times 10^6$ GeV \cite{El03}. For the pure
Einstein gravity five-dimensional model we obtain
\begin{equation}
E_K^{(3)}\approx \left(8.28-9.34\right)\times 10^{5} {\rm GeV}.
\end{equation}

From the point of view of multi-dimensional theories, the linear
model could describe the effects on the propagation of light in
Randall-Sundrum \cite{RaSu99} or Einstein-Yang-Mills \cite{Lo03},
\cite{HaCh04} type models. The quadratic model is specific for
quantum gravity effects, while the $\beta =3$ case could describe
the case of pure Einstein gravity in five dimensions. In all these
cases the delay of the $18$ GeV photon fixes the corresponding
energy scales.

The remarkable concordance between the linear quantum
gravity/extra-dimensional energy scale obtained from the present
study of the time delay of the $18$ GeV photon in GRB 940217 and
from the independent study of the BATSE and OSSE data \cite{El03}
strongly suggest that this time delay could be the signature of
the extra-dimensions or/and quantum gravitational effects.

Since the high energy photon in GRB 940217 is time delayed with
respect to the low energy photons, it follows that in the
five-dimensional line element $\varepsilon =-1$ and there is no
reason to consider super-luminal speeds for gamma-ray photons.
This result can also give some insights in the geometrical
structure of the fifth-dimensional Universe.

There are also some general arguments which suggest that the
signature of the five-dimensional metric corresponds to the choice
$\varepsilon =-1$. The five-dimensional field equations define a
cosmological constant $\tilde{\Lambda}=-3\varepsilon /L^{2}$,
where $L$ is a universal length, which has been introduced for
dimensional consistency \cite{MaHa99}. Therefore a measurement of
the sign of the cosmological constant may thus also be a
determination of the metric signature of the higher-dimensional
world. Lower limits on the age of the Universe as well as recent
observational data on high-redshift supernovae favor a positive
$\tilde{\Lambda}$. An other constraint on $\varepsilon $ can be
obtained from the study of the dynamics of a particle in five
dimensions. In our model we have explicitly assumed that the fifth
dimension $x^{4}=v$ is functionally related to the inertial
rest mass of the particle $m$ via Eq. (\ref{mass}), generally giving $m=m(s)$%
, where $s$ is the five-dimensional interval. The latter
expression does not imply a violation of the usual condition
$g_{\alpha \beta }u^{\alpha }u^{\beta }=1$ for the four-velocities
$u^{\alpha }=dx^{\alpha }/ds$, because it is a normalization
condition on the velocities and not a coordinate condition on the
metric. Multiplying by $m^{2}$ gives $p_{\alpha }p^{\alpha
}=m^{2}$, with no restriction on $m=m(s)$. In other words, the
energy $E^{2}=m^{2}c^{4}u^{0}u_{0}$ and the three-momentum $%
p^{2}=m^{2}c^{2}\sum_{i=1}^{3}u_{i}u^{i}$ can still satisfy an
equation of the form $E^{2}-p^{2}c^{2}-M^{2}(v)c^{4}=0$, with
$M^{2}(v)=m^{2}(s)\left[ 1-p_{4}p^{4}\right]=
m^{2}(s)\left[1-\varepsilon \Phi ^{2}\left(
t,v\right)(dv/ds)^2\right]$, even if the mass varies in
space-time. In order that these equations describe real particles
(with non-negative masses) it is necessary that the
five-dimensional metric should have a signature corresponding to
$\varepsilon =-1$, thus making the fifth dimension spacelike, in
agreement with the conclusion obtained from the study of the time
delay of the high energy photon in GRB 940217, and the observed
sign of the cosmological constant.

An alternative way of observationally testing multi-dimensional
models is the study of the gravitational Cherenkov radiation
\cite{MoNe01}. In some theoretical models the gravity can
propagate (in a preferred frame) with a
maximum velocity $c_{g}$, which can differ from the speed of light $c$, $%
c_{g}\neq c$. If the speed of gravity is slower than the speed of
light, particles moving faster than the speed of gravity would
emit a ''gravi-Cherenkov radiation'', in analogy with the
Cherenkov radiation emitted by particles moving faster than light
in a medium. Accurate measurements of the speed of propagation and
polarization of the gravitational waves can constrain brane world
theories, in which gravity propagates in the fifth dimension only,
while the matter is confined to four dimensions. The index of
refraction $n$ for the gravitational Cherenkov effect can be
estimated as $n-1\leq \left[ m_{Pl}^{2}/\left( 0.1E\right)
^{3}t\right] ^{1/2}$, where $m_{Pl}$ is the Planck mass and $t$
the travel time of the particle (for example, protons in cosmic
rays), with energy $E$ \cite{MoNe01}. The existence of high energy
cosmic rays which have travelled from astronomical distances
without losing all their energy to the gravitational Cherenkov
radiation places strong bounds on the speed of gravitational waves
with very short wave lengths \cite{MoNe01}. Accurate
determinations of the speed of the gravitational waves emitted by
distant astrophysical objects would allow the study of the
anomalous dispersion relation for gravitational waves, which is a
consequence of the violation of the Poincare invariance in the
higher dimension space-times, thus giving a better insight into
the geometrical structure of the five-dimensional space-time.
Therefore the analysis of the results obtained by using different
observational techniques, like the study of the time delay of high
energy photons from gamma ray bursts and  the study of the
gravitational waves, can strongly restrict energy scale of
multi-dimensional and/or quantum gravity effects and the value of
the coefficient $\beta $ in the generalized dispersion relation
given by Eq. (\ref{eq7}).

Since in the case of GRB 940217 only one high energy photon has
been detected, much more tests must be performed in order to
prove, via astrophysical observations, the existence of
extra-dimensions and of quantum gravity effects. A single photon
can also be generated as a result of an unusual fluctuation.
Therefore the study of a much larger number of high energy photon
delay events is necessary to firmly establish the astrophysical
effects of extra-dimensions.

 There are several proposed, satellite or ground based GRB research
projects which can perform this task. The Gamma-ray Large Area
Space Telescope (GLAST), a high energy ($30$ MeV-$300$ GeV)
gamma-ray astronomy mission, is planned for launch at the end of
2006. GLAST should detect more than 200 GRBs per year, with
sensitivity to a few tens of GeV for a few bursts \citep{We03b}.
GLAST is expected to observe, within 1 - 2 years of observations,
a large number of delayed photons, with energies of the order of
$10^2$ GeV. Their detection could provide evidence for the energy
and distance-dependent photon velocity dispersion predicted by
quantum gravity and multi-dimensional models.

The possibility that the very high energy component of the
gamma-ray bursts might be delayed makes the detection with the
sensitive ground based atmospheric telescopes more feasible.
Although TeV detectors already exist, their field of view is very
narrow ($\sim 1^{\circ }$). The sensitivity that can be achieved
is, for the current Whipple 10 m telescope of the order of
$8\times 10^{-8}$ ergs/cm$^2$ for a duration of the burst of $1$
and $10$ seconds, and of $2.4\times 10^{-7}$ ergs/cm$^2$ for a 100
s duration burst \cite{We03b}. For the proposed VERITAS array of
telescopes the minimum fluence is $10^{-8}$ ergs/cm$^2$ for $1$
and $10$ second bursts and $3\times 10^{-8}$ ergs/cm$^2$ for $100$
s bursts. These data can be regarded as representative of the
present generation of operating telescopes and for those that will
come on-line the next few years (CANGOROO III, HESS, MAGIC and
VERITAS) \cite{We03b}. The coordination between satellite
observatories and ground based gamma ray observatories
observatories will further improve the accuracy and precision of
the observations of the time delay.

Therefore, the detection of the time delay between Tev and GeV/MeV
photons from GRBs could represent a new possibility for the study
and understanding of some fundamental aspects of the physical laws
governing our universe.

\section*{Acknowledgments}

This work is supported by the RGC grant number HKU 7014/04P of the
 Government of the Hong Kong SAR of China. We thank T.C.
Weekes for useful discussions and the anonymous referee for
suggestions which helped us to improve the manuscript.

\end{document}